# Magneto-Seebeck effect in spin-valve with in-plane thermal gradient

S. Jain[1, a)], D. D. Lam[2, b)], A. Bose[1, c)], H. Sharma[3, d)], V. R. Palkar[1, e)], C. V. Tomy[3, f)], Y. Suzuki[2, g)] and A. A. Tulapurkar[1, h)]

*[1]Department of Electrical Engineering, Indian Institute of Technology Bombay, Powai, Mumbai-400 076, India*

*[2]School of Engineering Science, Division of Materials Physics, Osaka University, D312, 1-3 Machikaneyama, Toyonaka, Osaka 560-8531, Japan*

*[3]Department of Physics, Indian Institute of Technology Bombay, Powai, Mumbai-400 076, India*

We present measurements of magneto-Seebeck effect on a spin valve with in-plane thermal gradient. We measured open circuit voltage and short circuit current by applying a temperature gradient across a spin valve stack, where one of the ferromagnetic layers is pinned. We found a clear hysteresis in these two quantities as a function of magnetic field. From these measurements, the magneto-Seebeck effect was found to be 0.82%.

[a)]  Corresponding author: Sourabh Jain; E-mail: sourabhjain@ee.iitb.ac.in; Tel: +91 9029284615; Fax: +91 22 2572 3707; Postal address: Sourabh Jain, Department of Electrical Engineering, IIT Bombay, Powai, Mumbai, Pin-400 076;

[b)]  Duong Duc Lam; E-mail: lam@spin.mp.es.osaka-u.ac.jp; Tel/ Fax: +81 6 6850 6223;

[c)]  Arnab Bose; E-mail: arnabbose@ee.iitb.ac.in; Tel: +91 7208513596; Fax: +91 22 2572 3707;

[d)]  Himanshu Sharma; E-mail**:** himanshusharma@phy.iitb.ac.in; Tel: +91 9930924472; Fax:  +91 22 2576 7552;

[e)]  Vaijayanti R. Palkar; E-mail: palkar@ee.iitb.ac.in; Tel: +91 9819212819; Fax: +91 22 2572 3707;

[f)]  Chakkalakal V. Tomy; E-mail: tomy@phy.iitb.ac.in; Tel: +91 22 2576-7574; Fax:  +91 22 2576 7552;

[g)]  Y. Suzuki; E-mail: suzuki-y@mp.es.osaka-u.ac.jp; Tel/ Fax: +81 6 6850 6223;

[h)]  Ashwin A. Tulapurkar; E-mail: ashwin@ee.iitb.ac.in; Tel: +91 22 2576 7405; Fax: +91 22 2572 3707;

Generation and manipulation of spin current is an active area of research. Novel ways of generating spin current by spin pumping[1-5], spin Hall effect[6-8], spin-dependent Seebeck effect[9], etc. have become a major focus of spintronics. In particular, extensive research is going on in the area of "spincaloritronics"[10] to study the interplay of spins and temperature gradient. Spin-dependent Seebeck effect can be used to produce spin current by applying temperature gradient across a ferromagnet[11-13]. It has been predicted that this spin current can be used to even switch the magnetization of a nano-magnet via the spin-transfer torque effect[14,15]. Spin-dependent Seebeck effect has been investigated in magnetic multi-layers[16-19] and recently in magnetic tunnel junction pillers[20-23]. The reciprocal effect, viz., spin-dependent Peltier effect has also been reported[24-26]. A novel effect called Spin Seebeck effect related to the spin pumping has also been studied in various systems[27]. Here we present our results of the Seebeck effect measurements on a spin valve with in-plane thermal gradient where one of the magnetic layers is pinned. It is well known that the current-in-plane giant magneto-resistance (GMR) is sensitive to the reflection and transmission of electrons across non-magnetic and ferromagnetic interface. Thus our experiments in this geometry can measure the interface contributions to the spin dependent Seebeck effect.

The schematic of the spin-valve stack is shown in Fig. 1(a). Fabrication of the stacks was performed on thermally oxidized Si-substrates (using an E-880S-M ultra high vacuum system) in Ar-ambient with a base vacuum of $1 \times 10^{-9}$ mbar followed by the annealing at 300 °C for 2 hours at 6 kOe in-plane magnetic fields. Antiferromagnetic IrMn (7 nm) is used as a pinning layer in this structure. Ta (5 nm)/Ru (5 nm) buffer is used to promote IrMn into a fcc crystal structure[28,29]. The top CoFe (2 nm) layer is the free layer while the bottom CoFe (2 nm) layer is a pinned layer due to the FM/AFM exchange coupling between CoFe and IrMn. The free and fixed layers are separated by Cu (5 nm) spacer layer. Fig. 1(b) shows Magneto-Optic Kerr Effect (MOKE) signal of the spin-valve stack which confirms the existence of the pinned layer with exchange bias field $H_{ex}$ ~ 1.6 kOe. Fabricated GMR stack with a width of 1.5 mm is cut along the direction of the magnetization of pinned layer (x-axis). Two contacts are made at a distance $L = 7$ mm apart on the Ru layer using silver paint, as indicated in Fig. 1(a), to measure the voltage difference. We mounted one side of the sample on the heat sink (end 1) which is maintained at room temperature, and the other side on a heater (end '2') which is used to create the required temperature difference $\Delta T$. In-plane external magnetic field is applied with an angle $\theta$ with the x-axis, which is taken as the magnetization direction of the pinned layer.

The magneto-resistance of the GMR stack was measured first by passing an in-plane current of 100 µA (with no temperature difference applied) and sweeping the magnetic field along the direction of magnetization of pinned layer. Fig. 2(a) shows the hysteresis curve, $R$ vs. $H$, from which we obtained the resistance of anti-parallel state $R_{AP}$ ~ 34.54 Ω and resistance of parallel state $R_P$ ~ 34.26 Ω. The giant magneto-resistance (GMR) effect defined as ($R_{AP} - R_P$)/$R_P$ comes out to be ~0.82%.

In the next experiment, to study the magneto-Seebeck effect, we created a temperature difference across the sample by using the heater mounted on one end of the sample and measured the voltage difference between the two ends of the sample using a nano-voltmeter. The voltage was measured as a function of magnetic field swept along the direction of magnetization of pinned layer for various values of $\Delta T$. The results obtained after subtracting a field independent background voltage (which depends on the value of $\Delta T$), are plotted in Fig. 2(b) for various values of $\Delta T$. A clear hysteresis in the voltage as a function of magnetic field can be seen in Fig. 2(b). Comparing this figure with the magneto-resistance data (Fig. 2(a)), we can see that the hysteresis behavior in Fig. 2(b) corresponds to the change in the direction of magnetization of the free layer: for parallel alignment of the magnetization of free and fixed layer, the voltage measured is smaller than the voltage measured for the anti-parallel alignment. Further, this difference in voltage increases with increasing $\Delta T$ as can be seen from Fig. 2(b). Thus the voltage measured can be written as $V(\Delta T) = V_0(\Delta T) + V_{spin}(\Delta T)$. $V_{spin}$ depends on the relative orientations of the free and the fixed layer magnetizations whereas $V_0$ is independent of them.

Assuming linear response regime, the current flowing through our device can be written as[22, 30]:

$$I = G_V \Delta V + G_T (-\Delta T) \tag{1}$$

where $G_V$ and $G_T$ denote the electrical conductance and the thermoelectric coefficient, respectively. $\Delta V$ and $\Delta T$ denote the voltage and the temperature difference across the sample, respectively. Thus, under the open circuit condition ($I = 0$), application of temperature difference results in a voltage difference across the sample: $\Delta V = Q \Delta T$, where $Q = G_T/G_V$ is the Seebeck coefficient. Since in our device, the various coefficients depend on the relative magnetization directions, we write them as:

$$\begin{aligned} G_T^{P,AP} &= G_T^0 \pm (\Delta G_T / 2), \\ G_V^{P,AP} &= G_V^0 \pm (\Delta G_V / 2) \end{aligned} \tag{2}$$

where the superscript '0' denotes the average value. Thus, the relative magnetization dependent Seebeck coefficient can be written as:

$$Q^{P,AP} \approx Q^0 \pm Q^0 \frac{\Delta G_T}{2G_T^0} \mp Q^0 \frac{\Delta G_V}{2G_V^0} \qquad (3)$$

where $Q^0 = G_T^0 / G_V^0$ is the average Seebeck coefficient. From this we get the following expression:

$$\frac{\Delta Q}{Q^0} = \frac{\Delta G_T}{G_T^0} - \frac{\Delta G_V}{G_V^0} \qquad (4)$$

where $\Delta Q$ is defined as $\Delta Q = Q^P - Q^{AP}$. The open circuit voltage measured in the experiment is given by $(Q - Q_c)\Delta T$, where $Q_c$ is the Seebeck coefficient of the silver paint/Au wire used for making the contacts. We have plotted $V^P$, $V^{AP}$ and $\Delta V_{spin} = V^P - V^{AP}$ as a function of temperature difference in Fig. 3. The voltages depend linearly on $\Delta T$, implying that we are operating in the linear response regime. The value of $\Delta Q$ (= $\Delta V_{spin}/\Delta T$) can be obtained from Fig. 3. (left-axis) as 10.5 nV/K. We have further verified the spin-dependent Seebeck effect through a complimentary measurement of short circuit current.

The short circuit current was measured as a function of magnetic field swept along the direction of magnetization of pinned layer for various values of $\Delta T$. The results obtained after subtracting a field independent background voltage for $\Delta T = 40$ K are plotted in Fig. 4. A clear hysteresis in the current with a width of about 14 nA can be seen. The short circuit current is given by $V/R$, where $V$ is the open circuit voltage and $R$ is the resistance of the sample (and the connecting wires). Writing $V$ as $V = V_0 + V_{spin}$, the difference in the short circuit current can be written as:

$$\Delta I = I^P - I^{AP} = \frac{\Delta R}{R_0^2}V_0 + \frac{\Delta V_{spin}}{R_0} \qquad (5)$$

where $\Delta R = R^{AP} - R^P$ and $R_0 = (R^P + R^{AP})/2$. Using the values of $V_0 = -8.75$ μV and $\Delta V_{spin} = 420$ nV at $\Delta T = 40$ K, $\Delta R = 28$ mΩ, $R_0 = 34.4$ Ω, we see that the dominant contribution to $\Delta I$ comes from the second term as 12.21 nA, which matches fairly well with the experimental result shown in Fig. 4.

We further measured the angular dependence of the magneto-resistance and the magneto-Seebeck effect. An in-plane magnetic field with constant magnitude of 200 Oe was applied and the direction of the magnetic field ($\theta$) was swept from 0° to 360°. The results are shown in Fig. 5. We can see that both the resistance and the $V_{spin}$ show $\cos\theta$ dependence on the angle, which means that the contribution of the anisotropic magneto-resistance (AMR) effect is negligible, presumably due to the small thickness of the free layer compared to the other layers in the sample. To further verify this, we deposited a stack without the pinned layer, i.e. Si/SiO$_2$/Ta (5 nm)/Ru (5 nm)/CoFe (2 nm)/Cu (5 nm)/Ru (5 nm). The resistance measured by applying in-plane magnetic field with constant magnitude of 200 Oe and sweeping the direction of the magnetic field 0° to 360° is shown in Fig. 6. We see that the AMR effect is quite small ~ 0.04%, which is 20 times smaller than GMR effect in the previous sample. Further, the Seebeck voltage was found to be independent of angle (not shown in the figure). The above experiments were repeated for a stack with thick ferromagnet, viz., (Si/SiO$_2$/CoFeB (32 nm)/Ru (5 nm)). The magneto-resistance and magneto-Seebeck voltage as a function of angle are shown in Fig. 7. A clear $\cos2\theta$ dependence on the angle can be seen which are comparable to the results reported in references[31,32].

The average Seebeck coefficient of the GMR sample $Q_0$, can be obtained from Fig. 3 and using the relation, $V_0 = (Q_0 - Q_c)\Delta T$ as 1.28 µV/K, where we have used $Q_c$ = 1.5 µV/K for Ag/Au contact. Thus the value of $\Delta Q/Q_0$ measured in our experiment is 0.82 %. From equation 4, we can see that $\Delta Q/Q_0$ has contributions from two terms. From the second term, the magneto-resistance ratio was found to be 0.82%. Thus, the first term in equation 4, $\Delta G_T^0 / G_T^0$ contributes negligibly as compare to the second term.

In summary, we have measured the magneto-Seebeck effect in a spin-valve stack with "heat current in-plane geometry". We found that the dominant contribution to the magneto-Seebeck coefficient arises from the magneto-resistance effect.

We would like to acknowledge financial support by Department of Electrical Engineering, IIT Bombay as well as financial support provided by the Department of Electronics and Information Technology, Government of India, through the Centre of Excellence in Nanoelectronics and IITB Naofabrication facility.

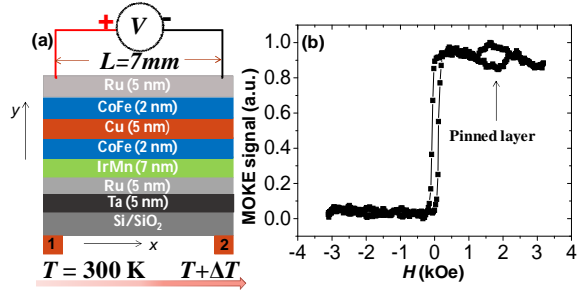

FIG. 1. (a) Schematic of our GMR device, placed across the thermal gradient. Cold end '1' is maintained at room temperature while hot end '2' is used for creating the temperature gradient $\Delta T$; (b) Magneto-Optic Kerr Effect (MOKE) signal of the GMR device is plotted against in-plane magnetic field $H$ at room temperature. Pinned layer is formed with exchange bias field of $H_{ex} \sim 1.6$ kOe.

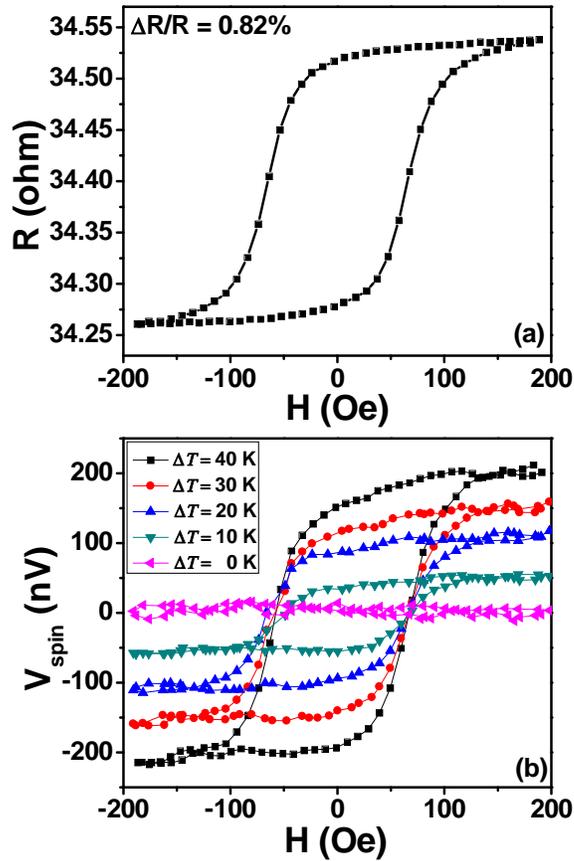

FIG. 2. (a) Electrical resistance $R$ and (b) Thermal voltage $V_{spin}$ are plotted as a function of in-plane switching magnetic field $H$ for the GMR device. The relative orientation of magnetic layers (parallel and anti-parallel

configuration of magnetization of free layer and pinned layer) leads to two different resistance states and causes GMR effect.

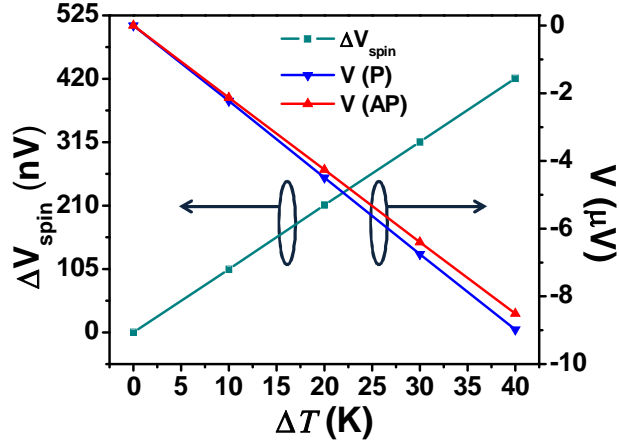

FIG. 3. Seebeck voltage $V$ is plotted against thermal gradient $\Delta T$, for anti-parallel (AP) and parallel (P) state of magnetization of free layer and pinned layer (see right-axis). The difference $\Delta V$ between $V$(AP) and $V$(P) is defined as the spin-dependent Seebeck voltage and plotted against $\Delta T$ (left-axis).

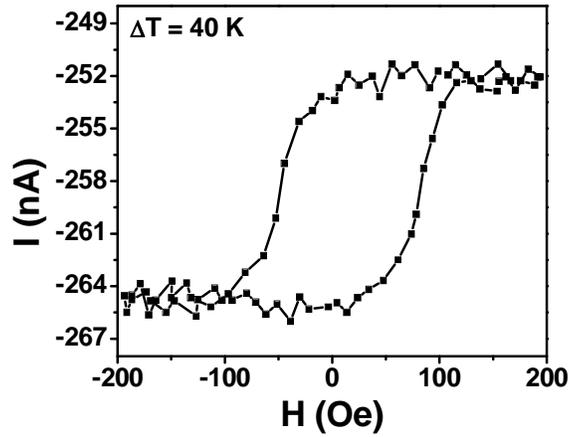

FIG. 4. Thermal-current $I$, plotted as a function of in-plane magnetic field $H$ at $\Delta T = 40$ K.

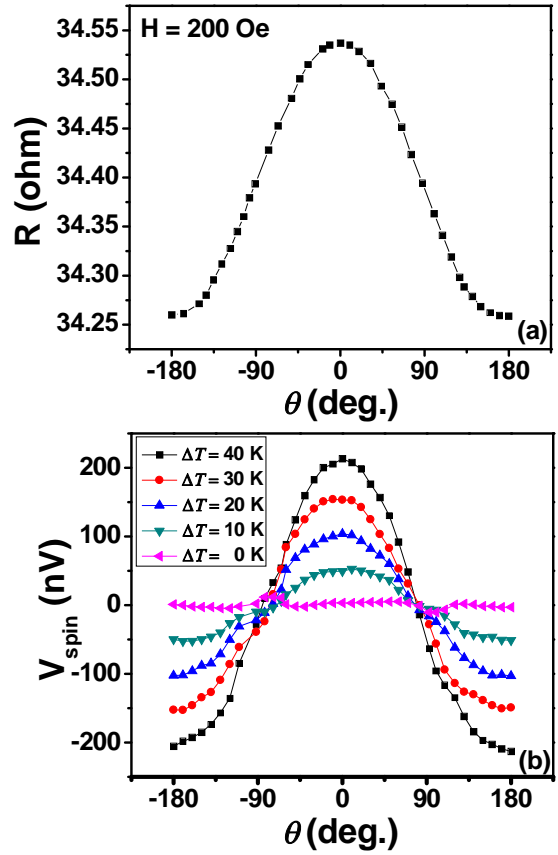

FIG. 5. (a) Electrical resistance $R$ and (b) Thermal voltage $V_{spin}$ in presence of in-plane magnetic field $H$ = 200 Oe, plotted as a function of $\theta$ which is the angle between the magnetization directions of free layer and pinned layer for the GMR device.

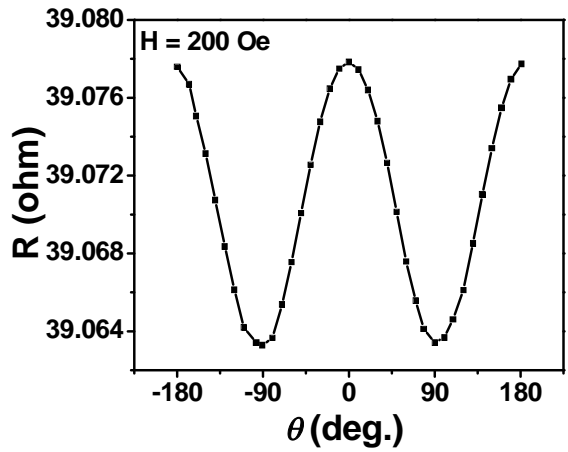

FIG. 6. (a) Electrical resistance $R$ versus $\theta$ in presence of in-plane magnetic field is plotted for 2 nm thick CoFe ferromagnet which is of the same thickness as in the GMR device.

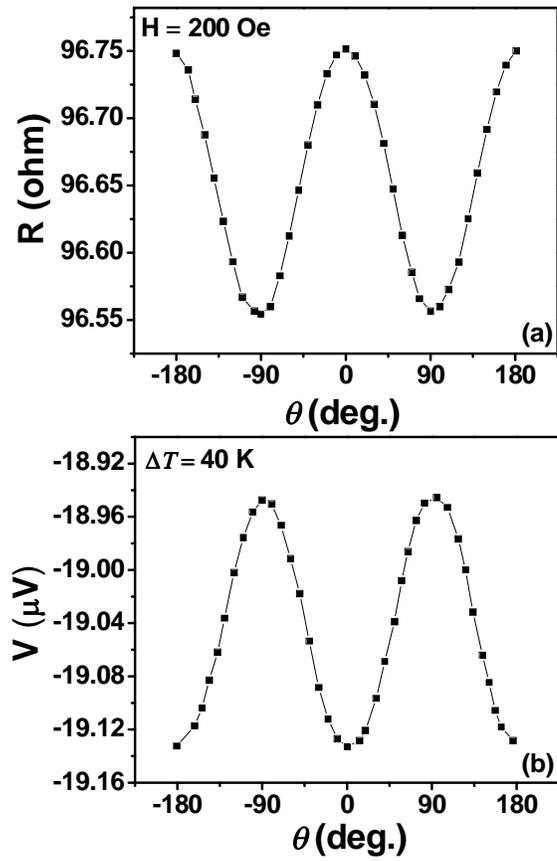

FIG. 7. (a) Electrical resistance $R$ and (b) Thermal voltage $V$ are plotted as a function of $\theta$ in presence of in-plane magnetic field for 32 nm thick CoFeB film.